\documentstyle[12pt,epsf]{article}

\def\thefootnote{\fnsymbol{footnote}}

\newcommand{\eq}{\begin{equation}}
\newcommand{\en}{\end{equation}}
\newcommand{\eqa}{\begin{eqnarray}}
\newcommand{\ena}{\end{eqnarray}}

\newcommand{\br}{\langle}
\newcommand{\kt}{\rangle}
\newcommand{\lb}{\lbrack}
\newcommand{\rb}{\rbrack}

\newcommand{\wh}{\widehat}

\newcommand{\lan}{\langle}
\newcommand{\ran}{\rangle}
\newcommand{\nonu}{\nonumber}

\newcommand{\dep}{\partial}

\newcommand{\NP}[1]{Nucl.\ Phys.\ {\bf #1}}
\newcommand{\PL}[1]{Phys.\ Lett.\ {\bf #1}}

\newcommand{\PR}[1]{Phys.\ Rev.\ {\bf #1}}

\begin{document}
\begin{titlepage}
\begin{flushright}
DFTT 4/2001\\
GEF-TH-04/2001\\
\end{flushright}
\begin{center}
{\Large\bf Correction induced by irrelevant operators}
\vskip 0.3cm
{\Large\bf  in the correlators   of the 2d Ising model}
\vskip 0.3cm
{\Large\bf   in a magnetic field.} 
\end{center}
\vskip 0.8cm
\centerline{
M. Caselle$^a$\footnote{e--mail: caselle@to.infn.it},
P. Grinza$^a$\footnote{e--mail: grinza@to.infn.it} and
N. Magnoli$^b$\footnote{e--mail: magnoli@ge.infn.it}}
 \vskip 0.6cm
 \centerline{\sl  $^a$ Dipartimento di Fisica
 Teorica dell'Universit\`a di Torino and}
 \centerline{\sl Istituto Nazionale di Fisica Nucleare, Sezione di Torino}
 \centerline{\sl via P.Giuria 1, I-10125 Torino, Italy}
 \vskip .2 cm
 \centerline{\sl  $^b$ Dipartimento di Fisica,
 Universit\`a di Genova and}
 \centerline{\sl Istituto Nazionale di Fisica Nucleare, Sezione di Genova}
 \centerline{\sl via Dodecaneso 33, I-16146 Genova, Italy}
 \vskip 0.6cm

\begin{abstract}
We investigate the presence of irrelevant operators in the
2d Ising model perturbed by a magnetic field, by studying
the corrections induced by these operators
 in the spin-spin correlator of the model. To this end we perform
a set of high precision simulations for the correlator both along the axes 
and along the diagonal of the lattice. By comparing the numerical results 
with the predictions of a perturbative expansion around the critical point 
we find unambiguous evidences of the presence of such irrelevant operators.  
It turns out that among the irrelevant operators the one
which gives the largest correction is the spin 4 operator $T^2 + \bar T^2$
which accounts for the breaking of the rotational invariance due to the lattice.
This result agrees with what was already known for the correlator evaluated
 exactly at the critical point and also with recent results
obtained in the case of the thermal perturbation of the model.

\end{abstract}
\end{titlepage}

\setcounter{footnote}{0}
\def\thefootnote{\arabic{footnote}}

\section{Introduction}
\label{s1}
 Since the seminal work of
 Belavin, Polyakov and Zamolodchikov~\cite{bpz} much progress has been done in
 the description of 2d statistical  models at the critical point.
 In particular in all the cases in which the 
 critical point theory can be identified
 with one of the so called ``minimal models''  a
 complete
 list of all the operators of the theories can be constructed. Moreover, it is
 possible to write differential equations 
 for the correlators and in some cases find their explicit expression in 
terms of
 special functions. Following these remarkable results, in 
 these last years a lot of efforts have  been devoted to extend them
  also outside the critical point. This can be achieved by perturbing the CFT
  with one (or more) of its relevant operators. 
  Much less is known in this case.  The most important result in this context is
  that for some particular choices of CFT's and
  relevant operators these
  perturbations give rise to integrable models~\cite{integrable,zam89}. 
 In these cases again we have a
 rather precise description of the theory. In particular it is possible to
 obtain the exact asymptotic expression for the large distance behavior of the
 correlators~\cite{form}. {}From this information several important results 
 (and in particular all the universal amplitude ratios) can be obtained.

  Similar efforts have also been devoted in trying to study vacuum expectation
values, amplitude ratios and correlators involving irrelevant operators. However
progress in this direction is limited by the lack of results (both numerical and
analytical) from actual
statistical mechanics realizations of the models.

 The only notable exception to this state of art is the 2d Ising model. In this
case, thanks to a set of remarkable results on the 2 and 4 point
correlators~\cite{ko49,fb67,w76}
 several interesting results on the contribution of
irrelevant operators have been obtained both in the case of the model at its
critical point~\cite{au-yang} and for $T\not=T_c$~\cite{n99,g2000} 
(i.e. for the theory perturbed by the energy operator). Very recently these
results have been used to study up to very high orders the contribution of
irrelevant operators to the magnetic susceptibility for 
$T\not=T_c$~\cite{g2000}. In this paper we shall 
try to extend this analysis to the case
of the magnetic perturbation of the Ising model. In particular, we shall look at
the contributions due to irrelevant operators to the spin-spin correlator in
presence of an external magnetic field. 
In particular we shall mainly concentrate on the
energy momentum tensor $T\bar T$, and the  
operator $T^2 +\bar T^2$ (which appear as a consequence of the breaking
due to the lattice of the full rotational symmetry) which are the 
 most important (i.e. those with smallest (in modulus) 
renormalization group eigenvalue) among the irrelevant operators. 

 In this case there is
no exact expression for the correlators or for the free energy of the model
 and the only way
that we have to judge of the validity of the CFT predictions is to
 compare them with the results of numerical simulations. Notwithstanding this
 there is a number of good 
reasons to choose exactly this model and this observable.
 Let us look at them in detail.
\begin{itemize}
\item
  The Ising model in a magnetic field is known to be
exactly integrable~\cite{zam89}, thus, even if it is
not exactly solvable,  a lot of important informations
can  all the same be obtained, in particular 
 very precise large distance expansions exist
for the correlators and the vacuum expectation values of the relevant operators
are known in the continuum limit~\cite{smatrix}.
\item
The model  is an optimal choice 
from the point of view of numerical simulations,
since very fast and efficient algorithms exist
 to study it.

\item
The model can also be realized as a particular case of the so called 
 IRF (Interaction Round a Face) models~\cite{E8a}. In this framework several
 interesting results can be obtained on the spectrum of the model~\cite{E8b}.

\item
By looking at the spin-spin correlator instead of the susceptibility we may
study the
 effects of the
 non zero spin operators which appear as a consequence of the breaking
of the rotational symmetry due to the lattice (see below for a precise
definition). These operators also appear in rotational invariant quantities like
the susceptibility, but only at the second order~\cite{chpv01,beta},
 i.e. at such an high power of
the perturbing constant that they can be observed only if the exact solution of
the model is known (as in the papers~\cite{n99,g2000}) and cannot be seen if one
can only resort to numerical simulations. 

\item
 A very interesting result of~\cite{g2000} (which had already been anticipated
 twenty years ago by Aharony and Fisher~\cite{af} and also confirmed
 in~\cite{gc}) is that in the thermal perturbation
of the Ising model the energy momentum tensor $T\bar T$ seems to be absent.
The first correction which involves irrelevant operators is thus due to the
spin 4
 operator $T^2 + \bar T^2$. This result should also hold at
 the critical point where it can be independently observed by using
 transfer matrix methods~\cite{chpv01}. However, apparently, it disagrees
 with the explicit form of the correlator at the critical point (which is
 exactly known on the lattice) which contains a scalar correction which can only
 be interpreted as due to the $T\bar T$ term. One of the aims of the present
 paper is  to understand this puzzle.

\item
If one is interested in studying irrelevant
operators it is mandatory to look at the short distance behaviour of the
correlator. In this respect the natural framework in which one must operate is
the so called IRS expansion~\cite{gm1,gm2}. This approach has been
recently discussed in great detail~\cite{cgm00} exactly in the 
case of the Ising model in a magnetic field in which we are interested here. 
It is only by using the results of~\cite{cgm00} as input of our analysis
that  we shall be able to reach the high level of
precision which is needed in order to observe the very small corrections which
are the signatures of the irrelevant operators.

\end{itemize}

This paper is organized as follows. In sect.2 we shall 
briefly summarize some known results on the 2d
Ising model in a magnetic field. In sect.3 we shall discuss the most important
contributions due to the irrelevant operators to the spin-spin correlators 
and shall evaluate their magnitude and behaviour.
In sect.4 we shall present the numerical
simulations that we have performed and finally in sect.5 we shall discuss the 
comparison between
numerical results and theoretical predictions. Sect. 6 is devoted to some
concluding remarks.

\section{The 2d Ising model in a magnetic field.}
\label{s2}
We shall be interested in the following in the Ising model defined on a 2d
square lattice of size $L$
 with periodic boundary conditions, in presence of an external
magnetic field $H$. The model is defined by the following
partition function:

\eq
Z=\sum_{\sigma_i=\pm1}e^{\beta(\sum_{\br i,j 
\kt}\sigma_i\sigma_j+H\sum_i\sigma_i)}
\en
where the notation $\br i,j \kt$ 
denotes nearest neighbour sites in the lattice.
In order to select only the magnetic perturbation, $\beta$ must be fixed to its
critical value:
 $$\beta=\beta_c=\frac12\log{(\sqrt{2}+1)}=0.4406868...$$

\noindent
 by defining  $h_l=\beta_c H$ we have
\eq
Z=\sum_{\sigma_i=\pm1}e^{\beta_c\sum_{\br i,j \kt 
}\sigma_i\sigma_j+h_l\sum_i\sigma_i}
\label{eq2}
\en
The magnetization $M(h)$ is defined as usual:
\eq
M(h)\equiv\frac1N\frac{\partial}
{\partial h_l}(\log~Z)\vert_{\beta=\beta_c}
= \br \frac1N\sum_i \sigma_i \kt .
\en
where $N\equiv L^2$ denotes the number of sites of the lattice.

 Eq.(\ref{eq2}) is the typical 
 partition function of a perturbed critical model.
With the choice $\beta=\beta_c$ the only  perturbing operator is
\eq
\sigma_l\equiv \frac1N\sum_i \sigma_i ~~~,
\en
We shall call in the following $\sigma_l$ as the spin operator (more precisely
the lattice discretization of the spin operator). Notice that
 the mean value of $\sigma_l$ coincides with $M(h)$:
\eq
\br \sigma_l \kt \equiv M(h)
\en
Our goal in this paper is to study the contribution of the 
 irrelevant operators to the spin-spin correlator. To this end we shall first
 study the model at the critical point (sect.\ref{s2.1}), we shall then switch
 on the magnetic field (see sect.\ref{s2.2})
 and discuss the modifications that it induces in the spin-spin correlator

\subsection{The Ising model at the critical point}
\label{s2.1}
\subsubsection{Operator content.}
\label{s2.1.1}
The Ising model at the critical point is described 
 by the unitary minimal CFT with central charge
$c=1/2$~\cite{bpz}.
Its spectrum  can be divided into three conformal families characterized by
different transformation 
properties under the dual and $Z_2$ symmetries of the model. They
 are the identity, spin and energy families and are
 commonly denoted as  $[{I}],~[\sigma],~[\epsilon]$.
Each family contains a ``primary'' field (which gives the name to the entire
family) and an infinite tower of ``secondary'' 
field (see below).
 The conformal weights of the primary operators are $h_{I}=0$,
 $h_\sigma=1/16$ and  $h_\epsilon=1/2$ respectively. 
Thus we see that in the Ising model the set of primary fields 
coincides with that of the relevant operators of the spectrum (remember that
the relationship between conformal
 weights and renormalization  group eigenvalues is: $y=2-2h$).
This is a peculiar feature of the Ising model only, and
 is not shared by any other minimal unitary model.
 Thus in this case the irrelevant operators are bound to be
secondary fields. Since in this paper we are particularly interested in the
irrelevant operators, let us study in more detail the structure of the three
conformal families.

\begin{itemize}

\item{\bf Secondary fields}

All the secondary fields are generated from the primary
 ones by applying the generators $L_{-i}$ and $\bar L_{-i}$ 
of the Virasoro algebra. In the following
we shall denote the most general irrelevant operator in the $[\sigma]$ family
(which are odd with respect to the $Z_2$ symmetry) with the notation
$\sigma_i$ and  the most
general operators belonging to the energy $[\epsilon]$ or to the identity $[I]$
families (which are $Z_2$ even) with $\epsilon_i$ 
and $\eta_i$ respectively.
 It can be shown that, by applying  
 a generator of index $k$: $L_{-k}$ or $\bar L_{-k}$  to a field $\phi$ 
 (where $\phi={I},\epsilon,\sigma$ depending on the case),
 of conformal weight $h_\phi$, a  new operator of weight
 $h=h_\phi+k$ is obtained.
 In general any combination of $L_{-i}$ and $\bar L_{-i}$
 generators is allowed, and the conformal weight of the resulting operator will
 be shifted by the sum of the indices of the generators used to create it.
 If we denote with $n$ the sum of the indices of the generators of type $L_{-i}$
 and with $\bar n$ the sum of those of type $\bar L_{-i}$ the conformal 
 weight of the resulting operator will be $h_\phi+n+\bar n$.
 The corresponding RG eigenvalue will be $y=2-2h_\phi-n-\bar n$.

\item{\bf Nonzero spin states}

 The secondary fields may have a non zero
spin, which is given by the difference  $n-\bar n$. In general one is
interested in scalar quantities and hence in the subset of those irrelevant
operators which have $n=\bar n$. However on a square lattice the rotation 
group is broken to the finite subgroup $C_4$ (cyclic group of order four).
Accordingly, only spin $0,1,2,3$ are allowed on the lattice.
If an operator $\phi$ of the continuum theory has  spin $j\in {\bf N}$,
then its lattice discretization $\phi_l$ behaves as a 
spin $j~({\rm mod}~4)$
operator with respect to the $C_4$ subgroup. As a consequence all the operators
which in the continuum limit have spin $j=4N$ with $N$ non-negative 
integer can appear in the lattice discretization of a scalar operator.

\item{\bf Null vectors}

 Some of the secondary fields disappear from the spectrum due to the null vector
 conditions. This happens in particular for one of the two states at level 2 in
 the $\sigma$ and $\epsilon$ families and for the unique state at level 1 in
 the identity family. From each null state one can generate, by applying the
 Virasoro operators a whole family of null states hence at level 2 in the
 identity family there is only one surviving secondary field, which can be
 identified with the stress energy tensor $T$ (or $\bar T$).

\item{\bf Secondary fields generated by $L_{-1}$}

 Among all the secondary fields a particular role is played by those generated
 by the $L_{-1}$ Virasoro generator. $L_{-1}$ is the generator of 
 translations on the lattice and as a consequence it has zero eigenvalue on
 translational invariant observables.

\item{\bf Quasiprimary fields}

A quasiprimary field $|Q \ran $ is a secondary field which is not a null vector
(or a
 descendent of a null vector) and satisfies the equation
\eq
L_1|Q \ran =0
\label{t1}
\en
These fields play a  central role in our analysis since they are the only
possible  candidates to be irrelevant operators of the model.

By imposing eq.(\ref{t1}) it is easy to construct  the first
few quasiprimary operators for each conformal family. 
For our analysis however the two lowest ones are enough. They both belong to the
conformal family of the identity. Their expression in terms of Virasoro
generators is:
\eq
Q_2^{\bf 1}=L_{-2}|{\bf 1} \ran 
\en
\eq
Q_4^{\bf 1}=(L_{-2}^2-\frac35L_{-4})|{\bf 1} \ran 
\en
(we use the 
 notation $Q^{\eta}_n$ to denote the quasiprimary state at
level $n$ in the $\eta$ family).

From these fields we can construct two irrelevant operators, which both have
conformal weight $4$ and RG eigenvalue $-2$.
\begin{description}
\item{1]}
The first combination is  $Q_2^{\bf 1} \bar{Q_2^{\bf 1}}$ which has spin zero
and can be identified with the energy momentum tensor $T\bar T$.
\item{2]}
The second combination is   $Q_4^{\bf 1} + \bar{Q_4^{\bf 1}}$
 which has spin 4
and can be identified with the combination $T^2 + \bar T^2$.
\end{description}
This last contribution appears as a consequence of the breaking of the full
rotational invariance due to the lattice\footnote{Notice that  
if we would be interested in scalar quantities, 
this term would disappear even on the lattice  at the first
order and could contribute only at the second order. This is the case for
instance of the susceptibility recently discussed in~\cite{g2000}, 
in which this
operator gives a contribution only at order $t^4$ and not $t^2$. However in the
present case since we are interested in a correlator, which defines
a preferential direction on the lattice this term can contribute already at the
first order. This fact will be discussed in detail sect.~\ref{s2.1.5} below.}.

\end{itemize}

\subsubsection{Structure constants.}
\label{s2.1.2}

Once the operator content is 
known,
the only remaining information which is needed to completely
identify the theory are the OPE constants. The OPE algebra is defined as 
\eq
 \Phi_i(r) \Phi_j(0) =
          \sum_{ \{ k \}} C^{ \{ k \}}_{ij} (r) 
           \Phi_{\{ k \}}(0) 
\en
where with the notation $\{ k \}$ we mean that the sum runs over all the 
fields of the conformal family $[k]$. 
The structure functions $C^{k}_{ij} (r)$ are c-number functions of $r$ which
must be single valued in order to take into account locality. 
In the large $r$ limit they decay with  a power like behaviour
\eq
C^{k}_{ij} (r) \sim |r|^{-dim(C^{k}_{ij})}
\en
whose amplitude is given by
\eq
\hat C^{k}_{ij}~\equiv~\lim_{r\to\infty} C^{k}_{ij} (r)~|r|^{dim(C^{k}_{ij})} .
\label{u1}
\en

Several of these structure constants are zero for symmetry reasons. These
constraints are encoded in the so called ``fusion rule algebra'' which, in the
case of the Ising model is.
\eq
\begin{array}{lll}
\lb \epsilon \rb \lb \epsilon \rb \ & = & \ \lb \mathbf{1} \rb \\
\lb \sigma\rb\lb\epsilon\rb \ & = & \ \lb\sigma\rb       \\
\lb\sigma\rb\lb\sigma\rb \ & = &\ \lb\mathbf{1}\rb \ + \ \lb\epsilon\rb.
\end{array}
\en
By looking at the fusion rule algebra we can see by inspection which are the
non-vanishing structure constants.

The actual value of these constants depends on the normalization of the fields,
which can be chosen by fixing the long distance behaviour of,
for instance, the $\sigma\sigma$ and $\epsilon\epsilon$ correlators. In this
paper we follow the commonly adopted convention which is:
\eq
\br \sigma(x)\sigma(0) \kt =\frac{1}{\,\,|x|^{\frac{1}{4}}}\,\,,
\hspace{1cm}|x|\rightarrow
\infty
\label{uv}
\en
\eq
\br \epsilon(x)\epsilon(0)\kt=\frac{1}{\,\,|x|^{2}}\,\,,
\hspace{1cm}|x|\rightarrow
\infty.
\label{uve}
\en
With these conventions  
we have, for the structure constants among primary fields
\eq
\hat C^{\sigma}_{\sigma,\sigma}=\hat C^{\sigma}_{\epsilon,\epsilon}=
\hat C^{\epsilon}_{\epsilon,\sigma}=0
\label{f4}
\en
\eq
\hat C^{{\bf 1}}_{\sigma,\sigma}=\hat C^{\sigma}_{\sigma,{\bf 1}}=
\hat C^{{\bf 1}}_{\epsilon,\epsilon}=\hat C^{\epsilon}_{\epsilon,{\bf 1}}=1
\label{f4b}
\en
and
\eq
\hat C^{\sigma}_{\sigma,\epsilon}= 
\hat C^{\epsilon}_{\sigma,\sigma}=\frac12 .
\label{f4c}
\en

\subsubsection{Continuum versus lattice operators.}
\label{s2.1.3}

Our main interest in this paper is the spin-spin correlator on the lattice. 
This rises the question of the relationship between the lattice and the 
continuum definitions of the operator $\sigma$.
In the following we shall denote the
 lattice discretization of the operators  with the index
 $l$. Thus $\sigma$ denotes the continuum operator and $\sigma_l$ the lattice
 one. 

 In general the lattice operator
is the most general combination of continuum operators compatible with the
symmetry of the lattice one. If we are exactly at the critical point this
greatly simplifies the analysis, since only operators belonging to the $[\sigma]
$ family are allowed. Moreover (due to the peculiar null state structure of the
spin family) the first quasiprimary operator in the spin family
appears at a rather high level and can be neglected in our analysis. Thus as far
as we are interested only at the critical point the spin operators on
the  continuum and on the lattice are simply related by a normalization
constant which we shall call in the following $R_\sigma$.
 The simplest way to obtain $R_\sigma$
 is to look at the analogous of eq.(\ref{uv}) on the lattice.
This is a well known result~\cite{wu}, which we report here for completeness.
The large distance behaviour of the correlator at the critical point is
\eq
 \label{def2}
\br \sigma_i \sigma_j \kt_{h=0}\ = \ \frac{ R_{\sigma}^2}
{|r_{ij}|^{1/4}}
\en
where  $r_{ij}$ denotes the distance on the lattice between the sites $i$ and
$j$ and
\eq
R^2_{\sigma}=e^{3\xi'(-1)}2^{5/24}=0.70338...
\en
By comparing this result with eq.(\ref{uv}) we find
\eq
 \sigma_{l} \ = \ R_{\sigma}  \sigma= 0.83868...
   \sigma
\label{rss}
\en

\subsubsection{The lattice Hamiltonian at the critical point}
\label{s2.1.4}
The last step in order to relate the continuum and lattice theories at the
critical point is the construction of the lattice Hamiltonian (let us call it
$H_{lat}$) at the critical
point. As above,
 the lattice Hamiltonian will contain all the operators compatible
with the symmetries of the continuum one.
In this case all the operators belonging to the $[\sigma]$ family are excluded
due to the $Z_2$ symmetry. Also the operators belonging to the $[\epsilon]$
family are excluded for a more subtle reason. The Ising model (both on the
lattice and in the continuum) is invariant under duality transformations while
the operators belonging to the $[\epsilon]$
family change sign under duality, thus they also cannot appear in
$H_{lat}(t=0)$. Thus we expect
\eq
H_{lat}~=~H_{CFT}~ +~ u^0_i~\int d^2x\eta_i, ~~~~~~~~~\eta_i \in [{I}]~~~,
\label{2.0bis}
\en
where $H_{CFT}$ is the continuum Hamiltonian and the $u^0_i$ are constants.
As mentioned above we can keep in this expansion only the first two terms which
are respectively $T\bar T$ and $T^2+\bar T^2$. 

\subsubsection{The spin-spin correlator on the lattice 
at the critical point.}
\label{s2.1.5}

We have already presented above, in eq.(\ref{def2})
 the large distance expansion of the spin-spin
correlator on the lattice at the critical point.

Thanks to the exact results obtained  in ~\cite{ko49,fb67,w76} (see~\cite{ising}
for a review) we
have much more informations on this correlator at the critical point.
In order to discuss these results let us first 
introduce a more explicit notation for the spins. We
shall denote in the following 
with $\sigma_{M,N}$ the spin located in the site with lattice
coordinates $(M,N)$. By using the translational invariance of the correlator we
can always fix one of the two spin in the origin. Thus the most general
correlator can be written as:
$ \lan \sigma_{0,0} \sigma_{M,N} \ran $. A particular role will be 
played in the following by the
correlator along the diagonal: $ \lan \sigma_{0,0} \sigma_{N,N} \ran $
and the one along the axis $ \lan \sigma_{0,0} \sigma_{0,N} \ran $.

Among the various results on the critical correlators two are of particular
relevance for us:
\begin{itemize}
\item
Remarkably enough, exact expressions exist for the correlator both along the 
diagonal and along the axis for any value of $N$. These expressions 
 are rather cumbersome and we shall not report them here. They
 can be found in~\cite{ising}.

\item
An asymptotic expansion exists for large values of the separation $r_{ij}$. 
For a square lattice, 
the first three terms have
been explicitly evaluated in~\cite{au-yang}.

\end{itemize}

This expansion takes the following form (for further details see \cite{au-yang})
\eqa
 \log \lan \sigma_{0,0} \sigma_{M,N} \ran & = & \log A -
\frac{1}{4} \log r + A_1 (\theta ) r^{-2} + 
\nonu \\
& + & A_2 (\theta ) r^{-4} + A_3 (\theta ) r^{-6} + O(r^{-8})
\ena
where
\eqa
A_1 (\theta ) & = & 2^{-8} (-1 + 3 \cos 4 \theta )
\nonu \\
A_2 (\theta ) & = & 2^{-13} (5 + 36 \cos 4 \theta + 36 \cos 8 \theta)
\nonu \\
A_3 (\theta ) & = & 3^{-1} 2^{-19} (-524 - 324 \cos 4 \theta + 24732 \cos 8
\theta + 28884 \cos 12 \theta)
\label{relat}
\ena
and 
\eqa
r^2 & = & \frac{1}{2}  ~ (M^2+N^2)
\nonu \\
M & = & r  ~ \sqrt2 \sin \theta
\nonu \\
N & = & r ~ \sqrt2 \cos \theta~.
\label{relat2}
\ena
In the following we shall only be interested to the first correction since
higher order corrections are beyond the resolution of our data.
Hence we end up with
\eq
\lan \sigma_{0,0} \sigma_{M,N} \ran = \frac{A}{r^{1/4}} \left( 1+ 2^{-8} (-1 +
3 \cos 4 \theta ) ~  r^{-2} +   \dots  \right)
\label{corrauya}
\en
Remarkably enough this expansion perfectly agrees with what one  
finds by assuming the presence in the Hamiltonian of the model of the two
irrelevant operators discussed in sect.\ref{s2.1.1}. In fact both $ T \bar T$
and $T^2 + {\bar T}^2$ have RG eigenvalue -2, thus, if they are present, they
should contribute to
the correlator exactly with a term proportional to $1/R^2$.
Moreover we expect that the $T^2 + {\bar T}^2$ operator, which has spin 4,
should give a term proportional to
$\cos(4\theta)$ while  the scalar operator $ T \bar T$ should give a
contribution without $\theta$ dependence. This is exactly the pattern that we
find in eq.(\ref{corrauya}) (see for instance pag. 218 of \cite{cardy} for a 
detailed discussion of this point).
However this remarkable agreement also rises a non trivial problem.
In fact the high precision analysis 
 of~\cite{g2000}, ~\cite{chpv01}
clearly exclude the presence in the lattice Hamiltonian of the $T\bar T$
operator (while they both confirm that the $T^2 + {\bar T}^2$ is indeed present)
It is thus not clear which could be the origin of the scalar term in 
 eq.(\ref{corrauya}).
A possible solution to this puzzle is to notice that the results of 
 eq.(\ref{corrauya}) and those of of~\cite{g2000}, ~\cite{chpv01} are
 obtained with two 
 different choices of the coordinates of the 2d plane.

In fact eq.(\ref{corrauya}) is written in terms of the 
``continuous" variable $r$ (which, is the one that we must choose if we want to
match the lattice results with the continuum limit ones).
Both the results of~\cite{g2000} and~\cite{chpv01} are instead obtained in
the ``lattice reference frame'' (which is the most natural variable on the
lattice) in which there is no $\sqrt2$ when comparing the distances along the
axes and along the diagonals. 
Hence, to compare the correction to the spin-spin correlation
function with the findings of 
 of~\cite{g2000}, ~\cite{chpv01}
 we must rewrite (\ref{corrauya})
it in terms of lattice coordinates. This can be easily performed 
by using the relations (\ref{relat2}). 
 
After some algebra we obtain
\begin{itemize}
\item{axis correlator
\eq
\label{inn1}
\lan \sigma_{0,0} \sigma_{0,N} \ran = \frac{A ~ 2^{1/8}}{N^{1/4}} \left( 1+
\frac{1}{64} ~  N^{-2} +   O (N^{-3})  \right)
\en
}
\item{diagonal correlator
\eq
\label{inn2}
\lan \sigma_{0,0} \sigma_{N,N} \ran = \frac{A}{N^{1/4}} \left( 1-
\frac{1}{64} ~  N^{-2} +   O (N^{-3})  \right).
\en
}
\end{itemize} 
These same expansion can also be obtained by directly
looking at the exact lattice results for the correlators (see for instance 
\cite{ising} where these $1/N^2$ terms are obtained in full detail).

Looking at eqs.(\ref{inn1}, \ref{inn2}) we see that with the lattice choice of
coordinates the
$1/N^2$ term exactly changes its sign as we move from the axis to the diagonal.
This is exactly what one would expect  
 for the contribution of a spin 4 operator, and thus it is apparent from
 eqs.(\ref{inn1}, \ref{inn2}) that 
no scalar correction appears at order $1/N^2$
in perfect agreement with the results of
 of~\cite{g2000}, ~\cite{chpv01}.

It is only the change of coordinates of eq.(\ref{relat2})
 which induces in the continuum
limit a scalar term (which we may well identify in this limit 
 with a $T\bar T$ type contribution).

%\vskip 1cm

\subsection{Adding the magnetic field.}
\label{s2.2}
\subsubsection{The continuum theory.}
\label{s2.2.1}
The continuum theory in presence of an external
 magnetic field is represented by the action:
\eq
{\cal A} = {\cal A}_0 + h \int d^2x \, \sigma(x) \,\,\,
\label{action}
\en

where ${\cal A}_0$ is the action of the conformal field theory.

As a consequence of the applied magnetic field the structure functions 
acquire a $h$ dependence so that we have in general
\eq
 \Phi_i(r) \Phi_j(0)  =
          \sum_{ \{ k \}} C^{ \{ k \}}_{ij} (h,r) 
           \Phi_{ \{ k \}}(0) ~~~~.
\en

Also  the mean values of the $\sigma$ and $\epsilon$
operators acquire a dependence on $h$. 
Standard renormalization group arguments (for un updated and thorough
 review on renormalization group theory applied to 
 critical phenomena see~\cite{pv00}) allow one to relate this $h$ dependence
to the scaling dimensions of the operators of the theory and lead to the
following expressions:
\eq
\langle\sigma\rangle_h=A_\sigma h^{\frac{1}{15}}+ ...
\label{mag}
\en
\eq
\langle\epsilon\rangle_h=A_\epsilon h^{\frac{8}{15}}+...
\en
The exact value of the two constants $A_\sigma$ and $A_\epsilon$
can be found in~\cite{fateev}  and \cite{flzz} respectively
\eq
A_\sigma=\frac{2\,{\cal C}^2}{15\,(\sin\frac{2\pi}{3}+
\sin\frac{2\pi}{5}+\sin\frac{\pi}{15})}=1.27758227..\,,
\label{sigmah}
\en
with
\eq
{\cal C} \,=\,
 \frac{4 \sin\frac{\pi}{5} \Gamma\left(\frac{1}{5}\right)}
{\Gamma\left(\frac{2}{3}\right) \Gamma\left(\frac{8}{15}\right)}
\left(\frac{4 \pi^2 \Gamma\left(\frac{3}{4}\right)
\Gamma^2\left(\frac{13}{16}\right)}{\Gamma\left(\frac{1}{4}\right)
\Gamma^2\left(\frac{3}{16}\right)}\right)^{\frac{4}{5}}~~  ,
\en
 and
\eq
A_\epsilon=2.00314...~~~.
\en

\subsubsection{Continuum versus lattice operators.}
\label{s2.2.2}

In presence of a magnetic field also the fields belonging to the 
energy and identity families can appear in the relation between the
 lattice and
the continuum version of the spin operator.
The most general expression is
\eq
\sigma_l=f^\sigma_0(h_l)\sigma+ h_l f^\epsilon_0(h_l)
\epsilon + f^\sigma_i(h_l) \sigma_i + h_l f^\epsilon_i(h_l)
\epsilon_i + h_l f^I_i(h_l)\eta_i,~~~  i\in{\bf N}
\label{xconvs}
\en
where $f^\sigma_i(h_l)$ $f^\epsilon_i(h_l)$ and $f^I_i(h_l)$ 
are even functions of $h_l$. With the notation $\sigma_i$, $\epsilon_i$ and
$\eta_i$ we denote the secondary fields in the spin, energy and identity
families respectively.
This is the most general expression, however, as a matter of fact,
only the first  term is relevant for our purposes (all the higher terms give
negligible contributions) 

\eq
\sigma_l=R_\sigma\sigma+hp_1\epsilon
\label{n1}
\en
The determination of $p_1$ is rather non-trivial. We shall discuss it in the
appendix. It turns out that 
\eq
p_1\sim 0.0345
\en
It is also important to stress that the lattice and continuum values of the
magnetic field do not coincide but are related by 
\eq
h_l=R_h h
\en
with $R_h=1.1923..$
(see ~\cite{cgm00} for details).
\subsubsection{The lattice Hamiltonian.}
\label{s2.2.3}
As it happened 
for the conversion from $\sigma$ to $\sigma_l$ also in the construction of
the perturbed Hamiltonian on the lattice, several new operators belonging to the
energy and spin family must now be taken into account. However it turns out that
all the quasiprimary fields of these two families appear at rather high level and
can be neglected in the present analysis. Thus, as far as we are concerned,
 the only effect of switching on
the magnetic field is that the constants in front of the $T\bar T$ and $T^2
+\bar T^2$ terms in eq.(\ref{2.0bis}) acquire an $h$ dependence. For symmetry
reasons these must be even analytic functions of $h$.

\subsubsection{The spin-spin correlator.}
\label{s2.2.4}
The short distance behaviour of the spin-spin correlator in presence of a
magnetic field can be obtained by using the so called IRS approach. A detailed
discussion of this approach can be found in~\cite{gm1,gm2}
 the particular application to the Ising correlators are
discussed in~\cite{cgm00} to which we refer for details. We only list here the
results.
Setting $t\equiv|h|~|r|^{15/8}$ we have up to $O(t^2)$

\eq
\br \sigma(0)\sigma(r) \kt |r|^{1/4}
= B_{\sigma\sigma}^1 + B_{\sigma\sigma}^2
t^{8/15}
+ B_{\sigma\sigma}^3  t^{16/15}+O(t^2) \label{ss}
\label{gmc}
\en

with
\eqa
B^{1}_{\sigma \sigma} & = & \wh{C^{\mathbf 1}_{\sigma \sigma}}~~=~~1 \nonu \\
B^{2}_{\sigma \sigma} & = &  A_{\epsilon} \wh{C^{\epsilon}_{\sigma \sigma}}~~
=~~ 1.00157... \nonu \\
B^{3}_{\sigma \sigma} & = &  A_{\sigma} \wh{\dep_h C^\sigma_{\sigma \sigma}}~~
=~~ -0.51581... \nonu\\
\ena

This result holds in the continuum theory. By using the known conversion between
continuum and lattice units we can obtain the spin-spin correlator on the
lattice:

\eq
\br \sigma_l(0)\sigma_l(r) \kt |r_{ij}|^{1/4}
= B_{\sigma\sigma,l}^1 + B_{\sigma\sigma,l}^2
t_l^{8/15}
+ B_{\sigma\sigma,l}^3  t_l^{16/15}+O(t_l^2) \label{ssl}
\en

Where $r_{ij}$ is the distance between the two spins on the lattice, measured in
units of the lattice spacing, $t_l$ is defined as  
$t_l\equiv|h_l|~|r_{ij}|^{15/8}$ and
\eqa
B^{1}_{\sigma \sigma , l} & = & 0.703384...\nonu \\
B^{2}_{\sigma \sigma , l} & = & 0.641409...\nonu \\
B^{3}_{\sigma \sigma , l} & = & -0.300749... \nonu\\
\ena

In the next section we shall see which contributions must be added to this
expression as a consequence of the irrelevant fields.

\section{Contribution from irrelevant operators to the spin-spin correlator 
in presence of a magnetic field.}
\label{s3}

There are two main sources of contributions due to 
 irrelevant operator to eq.(\ref{ssl}) above. They have different origin and
 contribute in a different way to the spin-spin correlator. 
The first correction can be understood as a perturbative contribution to the
spin-spin correlator due to irrelevant operators.  We shall discuss it in 
 subsec. 3.1.
The second correction arises from the new terms in the relation between 
$ \sigma _l $ and $\sigma $, due to the presence of a magnetic field.
We shall discuss it in 
 subsec. 3.2.

\subsection{Irrelevant operators in the perturbed Hamiltonian.}
\label{s3.1}
The contributions to the scaling function due to the (possible) presence
 of additional irrelevant 
operators in the lattice Hamiltonian
 can be studied by means of standard perturbative expansions
 around the CFT fixed point. \\
In order to address the problem in presence of an external magnetic field
 we must consider the CFT pertaining 
to the critical Ising model as perturbed by a mixed relevant/irrelevant 
perturbation. Hence there will appear correction terms to the scaling 
behaviour which both depend on the external magnetic field and show a
non-trivial (i.e. non scaling) dependence on the spin-spin distance $r_{ij}$.
Moreover if the irrelevant operator in which we are interested 
 breaks the 
rotational invariance (and this is the case for instance of the 
$(T^2 + \bar T^2)$ operator discussed above) we must expect a dependence on the
angle $\theta$ between the principal axes of the lattice and the direction of
the correlator. General symmetry arguments and dimensional analysis strongly 
constrain these terms. If we keep in our analysis only the first two irrelevant
operators $T\bar T$ and $(T^2 + \bar T^2)$ the most general expression for the
 scaling function turns out to be:  
\eqa
\br \sigma_l(0)\sigma_l(r) \kt |r_{ij}|^{1/4}
& = & B_{\sigma\sigma,
l}^1\left(1 + \frac{-1+a_1h^2}{|r_{ij}|^{2}}+\frac{(3+a_2h^2)\cos(4\theta)}
{|r_{ij}|^{2}}\right)  \nonu \\
& + & B_{\sigma\sigma,l}^2 \left(1 + \frac{b(\theta)}{|r_{ij}|^{2}}\right)
t_l^{8/15}  \nonu \\
& + & B_{\sigma\sigma,l}^3 \left(1 + \frac{c(\theta)}{|r_{ij}|^{2}}\right)
 t_l^{16/15}+O(t_l^2)
\label{ss2}
\ena
where $a_1$, $a_2$ are unknown constants and   $b (\theta)$, 
$c (\theta)$ are unknown functions of the form
\eq
b(\theta) = b_1 + b_{2}~cos(4 \theta), 
\label{nm1}
\en
\eq
c(\theta) = c_1 + c_{2}~cos(4 \theta), 
\label{nm2}
\en
in the angular variable $\theta$. \\
Let us analyze the above corrections in detail.
\begin{itemize}
\item{ Correction to $B_{\sigma\sigma,l}^1$:} 
 
At the lowest order in the irrelevant coupling, we can identify a scalar 
correction proportional to $|r_{ij}|^{-2}$ and a spin 4 correction proportional
 to $\cos (4 \theta) ~ |r_{ij}|^{-2}$. Both these terms acquire a $h$ 
dependence due to the magnetic relevant perturbation. The most general
expression, compatible with the symmetries of the model is:
\eq
B_{\sigma\sigma,
l}^1\left(1 + \frac{P_1 (h)}{|r_{ij}|^{2}}+\frac{P_2 (h)\cos(4\theta)}
{|r_{ij}|^{2}}\right)
\en  
where $P_1(h)$ and $P_2(h)$ are even polynomials in the magnetic field $h$
which in the $h \to 0$ limit must agree with the asymptotic expansion reported
in eq.(\ref{corrauya}). Expanding the polynomials up to the first term 
 in $h$ we find
\eqa
P_1(h) & = & -1 + a_1 h^2
\nonu \\
P_2(h) & = & 3 + a_2 h^2.
\label{pol}
\ena
from which the first term in eq.(\ref{ss2}) follows.
It is easy to see that the terms proportional to $a_1$ and $a_2$
 are highly suppressed due to the $h^2$ power.
They behave as the $O(t^2)$ that we have systematically neglected in the
previous section. As a matter of fact we have not been able to see these terms
in our numerical data and we shall neglect them in the following.

\item{Corrections to $B_{\sigma\sigma,l}^2$ and $B_{\sigma\sigma,l}^3$:}

Following the same line discussed above we find in this case:
\eqa
& B_{\sigma\sigma,l}^2 & \left(1 + \frac{b(\theta)}{|r_{ij}|^{2}}\right)
t_l^{8/15}  \nonu \\
& B_{\sigma\sigma,l}^3 & \left(1 + \frac{c(\theta)}{|r_{ij}|^{2}}\right)
 t_l^{16/15}
\label{cht}
\ena
where  $b(\theta)$ and $c(\theta)$ are the most general mixture of scalar and
spin 4 terms. They can be expanded in powers of $\cos(4\theta)$. keeping only
the first two orders in the expansion we end up with the expressions reported in 
 eq.s (\ref{nm1}) and (\ref{nm2}) where $b_1,b_2,c_1$ and $c_2$ are unknown
 constants.

While the term proportional to $c(\theta)$ 
 cannot be detected within the precision of
our data (we shall further discuss this point at the end of the next section),
 the   magnitude of the  one proportional to $b(\theta)$
 is definitely
larger than our numerical uncertainties. Thus we expect to be able to
observe such a correction.
\end{itemize}
\subsection{Irrelevant operators in $\sigma_l$.}
\label{s3.2}

We have seen above (see eq.(\ref{n1})) that in presence of a magnetic field the 
relation between lattice and continuum spin operator becomes more complicated
and a term proportional to $h\epsilon$ appears. Strictly speaking $\epsilon$
 is not an irrelevant operator, however this is only an accident, the following
 terms (those that we neglect) in the correspondence between $\sigma_l$ and
 $\sigma$ would indeed be irrelevant operators. Moreover,
 exactly as it happens for the irrelevant
 operators, also the $h\epsilon$ term
 gives a subleading contribution to the spin-spin correlator. For
 these reasons we have included also this correction in  this paper.

Inserting eq.
(\ref{n1}) in the spin-spin correlator we find
\eq
\br \sigma_l(0)\sigma_l(r) \kt =
R_\sigma^2\br \sigma(0)\sigma(r) \kt
+2hR_\sigma p_1\br \sigma(0)\epsilon(r) \kt
+h^2p_1^2 \br \epsilon(0)\epsilon(r) \kt
\label{uu1}
\en
Since we are interested in keeping only terms below
 $O(t^2)$ the last term in eq.(\ref{uu1}) can be neglected.
Using the known result (see~\cite{cgm00} for details)
\eq
 \br \sigma(0)\epsilon(r) \kt |r|^{9/8}
 = B_{\sigma\epsilon}^1  t^{1/15}
             +B_{\sigma\epsilon}^2 t+
B_{\sigma\epsilon}^3 t^{23/15} +
  O(t^{31/15}) \label{se}
\en
with
\eqa
B^{1}_{\sigma \epsilon} & = &  A_{\sigma} \wh{C^\sigma_{\sigma \epsilon}}~~=~~0.63879...
 \nonu\\
B^{2}_{\sigma \epsilon} & = & \wh{\dep_h C^{\mathbf 1}_{\sigma \epsilon}} 
~~=~~3.29627...\nonu\\
B^{3}_{\sigma \epsilon} & = &  A_{\epsilon} \wh{\dep_h C^\epsilon_{\sigma\epsilon}}
~~=~~-1.82085...\nonu
\ena
we can rewrite eq.(\ref{uu1}) as

\eqa
\br \sigma_l(0)\sigma_l(r) \kt  |r|^{1/4}
&=& R_\sigma^2 ( B_{\sigma\sigma}^1 + B_{\sigma\sigma}^2
t^{8/15}
+ B_{\sigma\sigma}^3  t^{16/15})\nonu \\
&+&2hR_\sigma p_1 |r|^{-7/8}  (B_{\sigma\epsilon}^1  t^{1/15}
             +B_{\sigma\epsilon}^2 t+
B_{\sigma\epsilon}^3 t^{23/15})
\label{uu2}
\ena
Only the first term in the second line of eq.(\ref{uu2}) gives a contribution
below $O(t^2)$ thus we end up with
\eqa
\br \sigma_l(0)\sigma_l(r) \kt  |r|^{1/4}
&=& R_\sigma^2 ( B_{\sigma\sigma}^1 + B_{\sigma\sigma}^2
t^{8/15}
+ B_{\sigma\sigma}^3  t^{16/15})\nonu \\
&+&2R_\sigma p_1 t |r|^{-22/8}  B_{\sigma\epsilon}^1  t^{1/15}
\ena
 
We see that the new term can be considered as a correction, proportional to $
\frac{1}{|r|^{22/8}}$ to the $t^{16/15}$ term  of 
eq.(\ref{ssl}):
\eqa
\br \sigma_l(0)\sigma_l(r) \kt |r_{ij}|^{1/4}
& = & B_{\sigma\sigma,l}^1 + B_{\sigma\sigma,l}^2
t_l^{8/15} \nonu \\
& + & (R_\sigma^2 B_{\sigma\sigma}^3 + 2R_\sigma 
p_1 |r|^{-22/8}  B_{\sigma\epsilon}^1 
)/R_h^{16/15} t_l^{16/15}
+O(t_l^2) \label{uu3}
\ena
Inserting the values of the various constants and of $p_1$ we end up with
\eq
\br \sigma_l(0)\sigma_l(r) \kt |r_{ij}|^{1/4}
= B_{\sigma\sigma,l}^1 + B_{\sigma\sigma,l}^2
t_l^{8/15}
+ (B_{\sigma\sigma,l}^2 + a_3 |r|^{-22/8}) t_l^{16/15}
+O(t_l^2) \label{uu4}
\en
with $a_3=0.0307...$

Unfortunately this new contribution is too small to give a reliable signature
within the precision of our data. In fact the amplitude of the correction
$a_3 |r|^{-22/8} t_l^{16/15}$  for distances greater than one lattice
spacing is always below (or at most of the order of) $10^{-4}$ and cannot
be disentangled from the other sources 
of corrections within the precision of our
data which ranges from $10^{-5}$ to $10^{-4}$. The situation is very similar to
that of the correction proportional to $c(\theta)$ discussed in the previous
section which in fact has the same $t$ dependence.

 However it is interesting to observe that both these corrections are
 almost in threshold to be observed 
and it is well possible that with the next generation of simulation they 
 could be detected. 
For this reason we included also their analysis in the present paper.
\section{The simulation}
\label{s4}
We studied the model with a set of Montecarlo simulations using a
  Swendsen-Wang type algorithm,
modified so as to take into account the presence of an external magnetic field.
We used the same program and simulation setting of ref.~\cite{cgm00} to which
we refer for 
 a detailed discussion of the performances of the algorithm and of finite size
 effects. We simulated the
 model for twelve
different values of the magnetic field, ranging from $h_l=4.4069\times10^{-4}$
to $h_l=8.8138\times10^{-3}$. 
 
For all the values of $h_l$ that we simulated, we studied the correlator:
$\langle \sigma(0)\sigma(r) \rangle$,
for $r=1,\cdots,10$ both in the diagonal direction and along the axes. This is
an important feature of the present analysis, since it is exactly by comparing
the correlator along the axes versus the diagonal one that we can extract the
spin of the perturbing operators.

For all
the simulations the lattice size was chosen to be $L=200$ (the analysis
of~\cite{cgm00} shows that for all the values of $h_l$ that we studied this is
enough to avoid finite size effects). Each two measures of the correlators were
separated by five SW sweeps. For each single correlator we performed
$4\times10^5$ measures. The values of $h_l$ that we studied are reported in
tab.\ref{tab1}
 where we also listed for completeness the corresponding values of the
correlation length $\xi$. This quantity is very important , since it defines in
a quantitative way what we mean as ``short distance behaviour''. Short distance
means ``smaller than the correlation length''. 
This is in fact the regime
in which we expect that the IRS analysis should hold and in which we may hope to
observe effects due to the irrelevant operators. $\xi$ is given (in lattice
units) by:
          
\eq
\label{xi}
\xi(h_l)=0.24935... h_l^{-\frac{8}{15}}
\en

\noindent
 (see~\cite{ch99} for details on this equation and on the 
 continuum to lattice conversion of $\xi$).

\begin{table}[h]
\label{tab1}
\caption{\sl The values of $h_l$ (measured in units of $\beta_c$) that we
studied and the corresponding correlations lengths.}
\vskip 0.2cm
\begin{tabular}{|c|c|}
\hline
$h_l$ &$\xi$    \\
\hline
$0.001$ & 15.4 \\
$0.002$ & 10.6 \\
$0.003$ & 8.5 \\
$0.004$ & 7.3 \\
$0.005$ & 6.5 \\
$0.006$ & 5.9 \\
$0.007$ & 5.4 \\
$0.008$ & 5.1 \\
$0.009$ & 4.8 \\
$0.01$ & 4.5 \\
$0.015$ & 3.6 \\
$0.02$ & 3.1 \\
\hline
\end{tabular}
\end{table}

\section{Comparison between predictions and numerical results.}
\label{s5}
Let us start this section
 by writing the expression that we expect for the spin-spin
correlator in a magnetic field, according to the analysis performed in the
sections 2 and 3 above. We have:
\eqa
\br \sigma_l(0)\sigma_l(r) \kt |r_{ij}|^{1/4}
& = & 
\br \sigma_l(0)\sigma_l(r) \kt_{h=0} |r_{ij}|^{1/4}
+ \left( a_1+ \frac{a_2 (\theta)}{\vert  r_{ij} \vert^2}+ \dots \right)
t_l^{8/15}
\nonu \\
& + & \left( b_1+ \frac{b_2 (\theta)}{\vert r_{ij} \vert^2} + \dots \right)  
t_l^{16/15} + O(t_l^2)
\label{sp4}
\ena 
where $\br \sigma_l(0)\sigma_l(r) \kt_{h=0} |r_{ij}|^{1/4}$ is given by 
eq.(\ref{corrauya});
$a_1 \equiv B^{2}_{\sigma \sigma, l}$ and $b_1 \equiv B^{3}_{\sigma \sigma, 
l}$, while $a_2 (\theta)$ and $b_2 (\theta)$ are in principle unknown 
functions of the angular variable $\theta$. \\
In this section we shall compare this expression with our numerical estimates
for
the correlators along the diagonal and the axis. This will allow us to obtain a
rather precise  estimate of the function 
$a_2(\theta)$ in the two cases $\theta=0$ and $\theta= \pi /4$.
With these two informations we shall also be able to study the spin content of
the operators which generate this contribution.\\
In the following we shall confine ourselves to the study of 
$a_2(\theta)$ since 
precision of our data is not 
enough to extend our analysis also to the function $b_2 (\theta)$.\\
We performed our analysis in three steps.
\begin{description}
\item{1]}
As a first step we constructed the combinations
\eq
G^{\Delta} (r,h_l) 
 \equiv \br \sigma_l(0)\sigma_l(r) \kt 
-\br \sigma_l(0)\sigma_l(r) \kt_{h=0} 
\en
using the fact that the critical point correlator is exactly known.
Thus we have:
\eqa
G^{\Delta} (r,h_l) 
|r_{ij}|^{1/4} & = &  (a_1+ \frac{a_2 (\theta)}{\vert  r_{ij} \vert^2}+ \dots)
t_l^{8/15}
\nonu \\
& + & (b_1+ \frac{b_2 (\theta)}{\vert r_{ij} \vert^2} + \dots) 
 t_l^{16/15} + O(t_l^2).
\label{inn3}
\ena 

\item{2]}
Then we study, at fixed value of $r$, the $h_l$ dependence of 
$G^{\Delta} (r,h_l)$. According to eq.(\ref{inn3}) above
we should choose as fitting function  
\eq
G^{\Delta} ( {\bar r} ,h_l) = A({\bar r}) h_l^{8/15} + B({\bar r}) h_l^{16/15}
\en 
(the notation $\bar r$ indicates that we are fitting at a fixed 
value of $\vert r_{ij} \vert$). \\
However keeping only the first two terms in the scaling function of the
correlator is a too rough approximation. Within the range of values of $h_l$
that we studied and with the precision that we obtained, this choice in general
led to very high  $\chi^2$ values. Fortunately, thanks to the IRS analysis
we know the functional form of the next to leading
 orders in the $h_l$ expansion of
the correlator. It turns out that adding one more term is enough 
for the correlators ranging from 1 to 
7 lattice spacings both along the axis and the diagonal directions,
while for the remaining correlators we have to go up to the fourth term in 
the expansion.
The general form of the scaling function at this order turns out to be:
\eq
G^{\Delta} ( {\bar r} ,h_l) = 
A({\bar r}) h_l^{8/15} + B({\bar r}) h_l^{16/15} +
C({\bar r}) h_l^{30/15} + D({\bar r}) h_l^{32/15}
\en 
(see~\cite{cgm00} for a discussion of this scaling function and the origin of
the two new exponents $h_l^{30/15}$ and $h_l^{32/15}$).
In this way for all the values of $r$ we found good confidence levels. 
The values of $A(r)$ that we obtained in this way
 can be found in table \ref{tab2}.
Notice that the data in each fit are completely uncorrelated since they belong
to different simulations. 
\item{3]}
As a last step we address the $r$ dependence of the function $A(r)$.
According to eq.(\ref{inn3})  we expect the following behaviour:
\eq
\frac{A (r)}{r^{3/4}} = a_1+ \frac{a_2(\theta)}{r^2}   
\en

Fixing $a_1$ to its known IRS value we end up with a one parameter fit which 
in both cases of $\theta=0$ and $\theta=\pi /4$ has a very good confidence 
level. 
Our final results are:

\begin{itemize}
\item{axis correlator 
\eq
a_2 (0) \ = \ (-0.062 \pm 0.004)
\en
}
\item{diagonal correlator 
\eq
a_2 (\pi /4 ) \ = \ (-0.014 \pm 0.007)
\en
}
\end{itemize}
\end{description}
These results, and in particular the fact that we have different corrections in
the two directions, unambiguously show that the contribution is at least 
partially due to a spin 4 operator. The natural candidate for this role is 
again the $T^2+\bar T^2$ operator. Our result also suggests that a scalar 
operator is present in the game. We see two possible reasons for this
contribution. The first is that, since
 we are
working in the IRS framework we are compelled to use the ``continuum limit
reference frame''. As  we have seen in sect.\ref{s2.1.5} this choice induces a
$T\bar T$ term and consequently a scalar type correction. The second is
that one cannot exclude the presence in the magnetic scaling field of a
(rotationally invariant)
momentum dependent term, which would manifest itself exactly as a scalar 
correction of
the type that we observe\footnote{We thank A.Pelissetto for pointing to 
us this interesting possibility.}.
 This non trivial possibility has been recently
discussed in~\cite{beta} (see the note at pag.8161) to explain the corrections
of order $t^2$ (where $t$ is the reduced temperature) in the second moment of
the spin spin correlator. Most probably both mechanisms are at work in the
present case. We cannot distinguish between them within the precision of our
data.

\begin{table}[h]
\label{tab2}
\caption{\sl The lattice spacing $N$ is intended along the axis for $A^{axis}$ and along the diagonal for $A^{diag}$.}
\vskip 0.2cm
\begin{tabular}{|c|c|c|}
\hline
$ N $ & $A^{axis}(N)$ & $A^{diag}(N)$    \\
\hline
$1$ & $ (0.581 \pm 0.003) $ & $(0.823 \pm 0.003)$     \\
$2$ & $ (1.048 \pm 0.004) $ & $(1.396 \pm 0.004)$       \\
$3$ & $ (1.446 \pm 0.005)$ & $(1.894 \pm 0.005) $   \\
$4$ & $ (1.802 \pm 0.005)$ & $(2.354 \pm 0.007) $   \\
$5$ & $ (2.134 \pm 0.006)$ & $(2.777 \pm 0.008)   $ \\
$6$ & $ (2.448 \pm 0.006)$ & $(3.19 \pm 0.01) $  \\
$7$ & $ (2.746 \pm 0.007)$ & $ (3.58 \pm 0.01) $  \\
$8$ & $ (3.039 \pm 0.009)$ & $  (3.96 \pm 0.01)$   \\
$9$ & $ (3.316 \pm 0.009)$ & $(4.32 \pm 0.03)  $   \\
$10$ & $ (3.58  \pm 0.01)$  & $(4.67  \pm 0.04)$     \\
\hline
\end{tabular}
\end{table}
\section{Concluding remarks}
\label{s6}
The role of the irrelevant operators in the two dimensional Ising model 
has attracted much interest in these last months due to the results on
the magnetic susceptibility at $h=0$ recently reported in~\cite{n99,g2000}.
While it is by now clear that  contributions due to irrelevant operators are
present in the free energy of the Ising model at $h=0$ nothing is known on the
case in which the magnetic perturbation of the critical model is chosen.
Moreover the characterization of these irrelevant operators and, possibly,
their identification with quasiprimary fields of the Ising CFT is still an open
problem.  In this paper we tried
to make some progress in this direction.
We studied the corrections induced by the presence
 of irrelevant operators in the spin-spin correlator of the 2d Ising model in
 presence of an external magnetic field. 

Our main results are:
\begin{itemize}
\item
The $1/r^2$ corrections which are present
 in the correlator at the critical point survive unchanged also if the magnetic
 field is switched on. The $h$ independent part of this contribution 
is much larger than our statistical errors and can
 be observed very precisely. It is completely due to the spin 4 irrelevant
 operator.

\item
Several new terms appear 
 when the magnetic field is switched on. By using standard perturbative methods
 one can evaluate the amplitude of the corrections that they induce in the 
spin-spin correlator. In general these terms are very small. However in one case
 the amplitude of the expected correction  turned out to be
 definitely larger than our statistical errors.
It is the case of the term  proportional to
\eq
\frac{1}{|r|^{1/4}}\left(\frac{t^{8/15}}{r^2}\right)~~~~.
\en 
for which  we could indeed
 observe and measure such correction  with good confidence. 
The comparison between the two values of this correction for the correlator
along the diagonal and the one along the axis shows that,
besides the expected contribution due to the spin-4~~ $(T^2 +\bar T^2)$
operator, there is also a scalar term which may have
 a twofold origin. It could be due the
$T\bar T$  operator, but it could  also be the signature
of a momentum dependent term in the magnetic
scaling field of the model.

\end{itemize}

Other contributions due to these or other
 irrelevant operators are for the moment beyond our
resolution, but it could be possible to observe them in the next generation of
simulations.

\newpage

\appendix{}
\section{Determination of $p_1$}
\renewcommand{\theequation}{A.\arabic{equation}}
\setcounter{equation}{0}

The simplest way to fix the value of $p_1$ is to look at the expectation value
$\br \sigma_l \kt$.
We expect:
\eq
\br \sigma_l \kt=R_\sigma \br \sigma \kt +p_1 \br \epsilon \kt
\en
from which we have (using the notations of \cite{ch99})
\eq
\br \sigma_l \kt=R_\sigma A_M h^{\frac{1}{15}}
+ p_1 A_E h^{\frac{23}{15}}
\en
from \cite{ch99} we see that
\eq
\br \sigma_l \kt=\frac{16}{15} A_f^l h^{\frac{1}{15}}
+\frac{38}{15} A_f^lA_{f,2}^l h^{\frac{23}{15}}
\en
Using the numerical values reported in~\cite{ch99}, i.e.
\eq
A_f^l=0.9927995....
\en
\eq
A_{f,2}^l\sim 0.021...
\en
we obtain
\eq
p_1\sim 0.0345
\en

\vskip 3cm

{\bf  Acknowledgements}
We thank M. Hasenbusch, A.~Pelissetto and E.~Vicari for useful
discussions and correspondence on the subject.

\end{document}